Bifacial weakness with paresthesias(BFP) secondary to trauma: a case report

Jingjing Chen[a,1], Xuxia Tang[a], Shuo Dai[a], Xiao He[a,*]

[a]Department of Otolaryngology, The First Affiliated Hospital of Zhejiang Chinese Medical University (Zhejiang Provincial Hospital of Chinese Medicine), Hangzhou, 310000, China

*Corresponding author

E-mail adress: 202011021512070@zcmu.edu.cn

Abstract:

This case details the diagnosis and treatment process of a patient with bilateral facial nerve palsy accompanied with limb sensory disturbance secondary to head trauma, who was ultimately diagnosed with Bifacial weakness with paresthesias (BFP) , a rare variant of Guillain-Barré Syndrome(GBS) . The patient underwent plasma exchange therapy and showed favorable recovery . In this article, for the first time we report a case of BFP secondary to trauma.

1. Introduction:

Guillain-Barré syndrome (GBS) is an acute immune-mediated polyneuropathy characterized by rapidly progressive symmetric limb weakness and diminished or absent tendon reflexes, often accompanied with sensory disturbances. Its pathogenesis is closely related to damage to the peripheral nerve myelin or axons caused by immune-mediated mechanisms. While classic GBS is well-known among neurologists, its variant forms present with complex and diverse manifestations, often leading to clinical misdiagnosis or missed diagnosis. Bifacial weakness with paresthesias (BFP) is a rare variant of GBS. When triggered by head trauma, the attribution of bilateral facial paralysis becomes particularly challenging, posing a significant diagnostic dilemma for first-contact physicians. Departments such as otolaryngology and emergency often serve the initial points of care for such patients. Subtle sensory abnormalities may be easily overlooked, leading to an initial diagnosis skewed toward traumatic facial nerve injury.

This paper aims to illustrate the clinical course and diagnostic challenges of BFP following trauma through a clinical case. We emphasize that, in cases of bilateral facial paralysis after trauma, clinicians should remain highly vigilant, detailed inquiries and examinations are essential.

2. Case Report:

A 47-year-old male patient was transferred to our department 7 days after a car accident with bilateral facial nerve palsy for 5 days. He also presented with hearing loss, decreased taste sensation, and sensory impairment in both lower limbs. The patient received dexamethasone intravenously for 4 days at at the first hospital, but there was no improvement in the facial paralysis. Physical examination revealed absence of forehead wrinkles, incomplete eye closure, flattened bilateral nasolabial folds, due to bilateral facial paralysis, the patient was unable to show the teeth(Fig.1) It was corresponding to House-Brackmann facial nerve grading scale grade V. MRI revealed a left temporal lobe hematoma, fluid accumulation in bilateral mastoid processes and left sphenoid sinus, CT scan showed bilateral temporal bone fractures without involvement of the facial nerve canal(Fig. 2). Initially, we still considered bilateral facial nerve paralysis due to temporal bone trauma and continued glucocorticoid therapy. The turning point occurred on the third day after the patient was transferred to our department, a neurologist observed diminished tendon reflexes and strongly suspected Guillain-Barré syndrome (GBS) during the consultation. To confirm the diagnosis, a cerebrospinal fluid (CSF) test was performed, along with an anti-ganglioside antibody test. The CSF results revealed protein-cytological dissociation. Based on the medical history and CSF test, the patient was diagnosed with BFP, a rare variant of GBS.

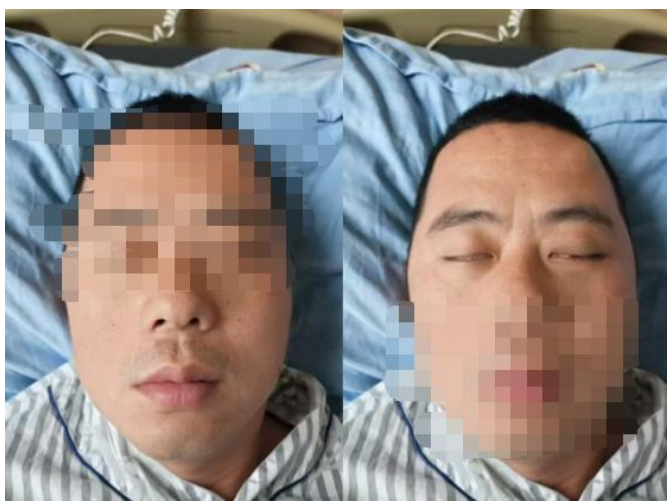
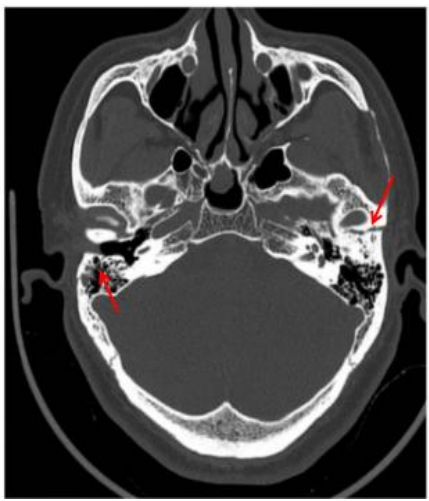
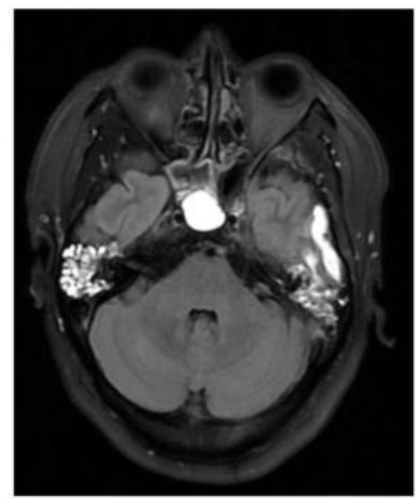

Fig.1. 5 days after trauma, the patient presented with incomplete eye closure, flattening of the nasolabial fold

Fig.2. CT scan shows bilateral temporal bone fractures and MRI reveals a left temporal lobe hematoma, fluid accumulation in bilateral mastoid processes and left sphenoid sinus

The patient received plasma exchange therapy. On the 22th day after the onset of facial paralysis, symptoms began to show signs of improvement. 2 months post-injury, there was essentially no liquid leakage while drinking, but recovery of eye closure, forehead wrinkles remained unsatisfactory(Figs, 3 ). 6 months post-injury revealed complete eye closure without scleral show, normal tooth display and forehead wrinkles, corresponding to House-Brackmann grade I(Figs. 4).

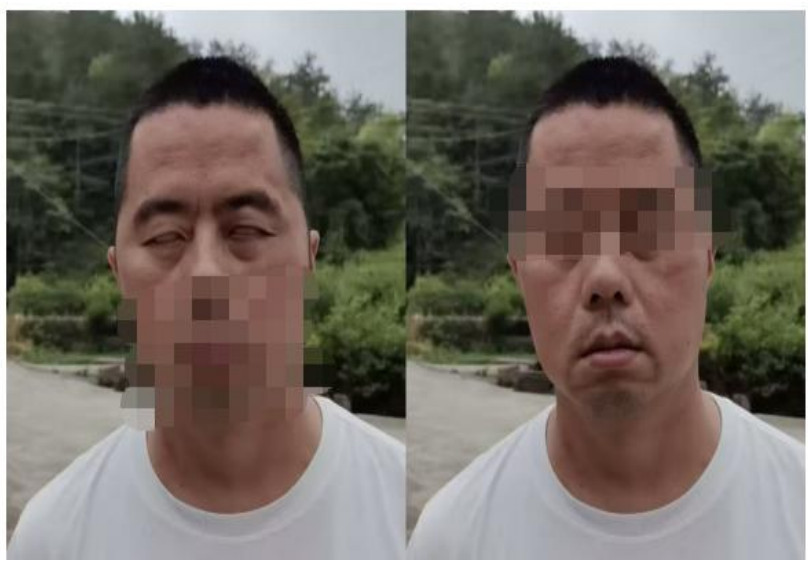
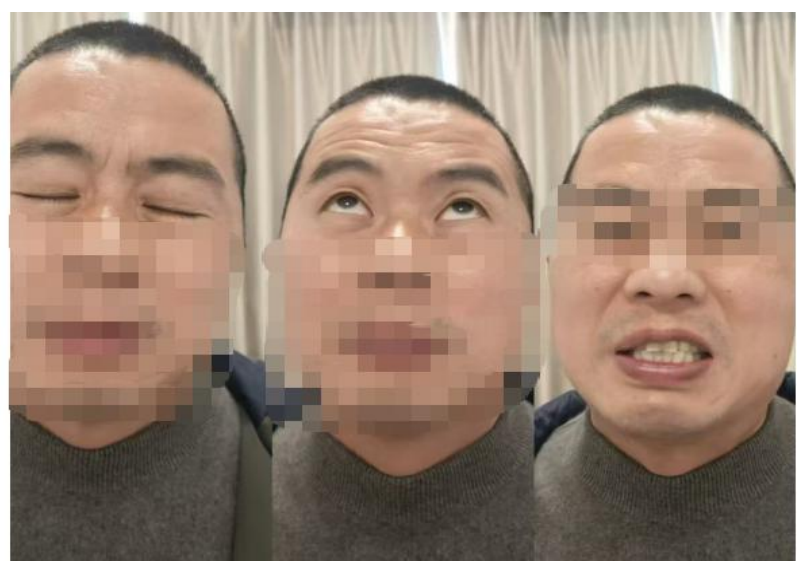

Fig. 3. 2 months post-injury, the patient presented improvement in the nasolabial fold flattening, but still incomplete eye closure

Fig. 4. 6 months post-injury, the patient showed good eye closure, normal forehead wrinkles and normal nasolabial folds

3. Discussion:

Bilateral facial nerve paralysis following trauma is rare and can be categorized into immediate and delayed onset. Immediate facial paralysis often indicates severe injury to facial nerve, like entrapment, laceration, contusion, crushing, or traction of the facial nerve. Delayed facial nerve paralysis typically occurs 2–21 days after the trauma and is more common than immediate onset, associated with edema, arterial or venous thromboses, or bleeding into the facial nerve bony canal that leads to external compression of the nerve[1]. Literature review reveals that cases of bilateral delayed facial paralysis after trauma are case reports. These case reports of bilateral facial nerve paralysis after trauma indicate that the injury is the direct cause of the facial paralysis. Abrahão NM, et al. performed a facial nerve decompression surgery on a patient with bilateral facial paralysis following trauma, with partial recovery of facial paralysis during follow-up[2]. Medha KK, et al. treated a patient with bilateral facial paralysis after trauma with corticosteroid and physiotherapy , showing satisfactory recovery after six months of follow-up[3]. Based on these previous reports, we initially treated this patient with glucocorticoid. Facial paralysis, decreased taste sensation and hearing loss, all of these can be attributed to the trauma, coupled with the confidence derived from previous literature, the lack of immediate efficacy did not lead us to suspect the diagnostic.

GBS is not a common etiology of acute-onset bilateral facial paralysis. Keane JR[4] analyzed the causes in 43 patients with bilateral facial paralysis and found that the most common condition is Bell's palsy (10 cases), followed by tumors (9 cases), and then GBS and infections (including tuberculosis, syphilis, and HIV). Neurosarcoidosis and neuroborreliosis also require differential diagnosis. This patient had no involvement of other organs such as the lungs, eyes, or lymph nodes, and the disease course was acute and monophasic, thus ruling out sarcoidosis. There was no history of rash or tick bite prior to onset, and the absence of lymphocytosis in the cerebrospinal fluid examination excludes Lyme disease. Systemic imaging and blood tests ruled out infectious diseases and tumors. Keane JR also reminds clinicians to pay attention to diminished or absent tendon reflexes, a physical examination finding suggestive of GBS.

GBS is typically triggered by infections, while post-traumatic GBS is mostly limited to case reports[5,6,7]. Helen Grote et al. encourage clinicians to register any cases of post-traumatic or post-surgical GBS with international registers to raise awareness[6]. This patient had no symptoms of respiratory or gastrointestinal infection within four weeks prior to onset, no recent vaccinations, and tested negative for HIV and syphilis. To rule out invisible damages caused by the trauma, chest and abdominal CT scans were performed, which also revealed no tumors. Therefore, we have reason to believe that the patient's GBS was trauma-induced. GBS has diverse subtypes, with BFP is a rare variant characterized by rapidly progressive bilateral facial weakness in the absence of other cranial nerve involvement, ataxia, or limb weakness. Many patients also experience distal limb paresthesia. In 1962, Charous and Saxe reported a case of isolated facial paralysis with CSF protein-cytological dissociation, without involvement of other cranial nerves or limbs. Symptoms resolved after three weeks, they were the first to associate bilateral facial paralysis with GBS[8]. Thirty years later, Ropper described four subtypes of GBS, including BFP[9]. In 2009, to determine the clinical features of BFP, Susuki K et al. [10] reviewed over 8,600 serological samples and medical histories of neurological diseases from their neuroimmunology laboratory, ultimately diagnosing 22 patients with BFP. All patients presented with CSF protein-cytological dissociation. Among them, 86% initially experienced limb numbness, indicating that subtle symptoms of generalized polyneuropathy (paresthesia) are a key feature of BFP. Mild distal limb paresthesia is a symptom easily overlooked by both clinicians and patients, and it was initially missed in our case as well until further questioning when detecting diminished tendon reflexes. In fact, limb paresthesia along with diminished or absent tendon reflexes are sufficient to distinguish BFP from most other causes of facial paralysis. Among the 22 patients, 82% had a history of preceding infections, indicating that infection remains the primary trigger, consistent with classic GBS. Our case, with no history of preceding infection, is the first time to report a case of BFP secondary to trauma. Eleven patients experienced dysgeusia. While classic GBS rarely involves taste abnormalities, dysgeusia is relatively common in Bell's palsy and facial paralysis caused by trauma. Five patients tested positive for anti-GM2 IgM antibodies, with no anti-ganglioside IgG antibodies detected. Our patient was positive for anti-GM2 IgM.

Intravenous Immunoglobulin Therapy (IVIG) and plasma exchange are standard treatments for GBS. In a study by Susuki K[10], six patients received IVIG, four underwent plasma exchange, and three received steroid therapy. Except for one patient who still had severe facial weakness at the 10th month, most patients had a favorable prognosis. Our patient underwent plasma exchange and showed good recovery. According to previous literature, it is difficult to conclude that IVIG and plasma exchange are superior to observation. Although BFP typically has a favorable prognosis, patients with classic GBS may initially present with neurological symptoms similar to BFP and later progress to quadriplegia[4]. Therefore, close monitoring symptoms of patients presenting with BFP is necessary , the standard treatments for GBS are reasonable.

4. Conclusion:

We report a case of BFP secondary to trauma, which has not been previously documented. We emphasize that when encountering bilateral facial nerve paralysis following trauma, clinicians should avoid prematurely attributing the condition solely to the injury. Instead, a comprehensive history inquiry and physical examination must be conducted to rule out underlying diseases.